\documentclass[twocolumn,prl,amsmath,amssymb]{revtex4}
\usepackage{graphicx}
\usepackage{dcolumn}
\usepackage{bm}

\begin{document}

\title{Raman Fingerprint of Charged Impurities in Graphene}
\author{C. Casiraghi$^1$, S. Pisana$^1$, K. S. Novoselov$^2$, A. K. Geim $^2$, A. C. Ferrari$^1$}\email{acf26@eng.cam.ac.uk}
\affiliation{$^1$Engineering Department, Cambridge University, Cambridge, UK\\
$^2$Department of Physics and Astronomy, Manchester University, Manchester, UK\\}

\begin{abstract}
We report strong variations in the Raman spectra for different single-layer graphene samples obtained by micromechanical cleavage, which reveals the presence of excess charges, even in the absence of intentional doping.
Doping concentrations up to $\sim$10$^{13}$cm$^{-2}$ are estimated
from the G peak shift and width, and the variation of both position and relative intensity of the second order 2D peak. Asymmetric G peaks indicate charge
inhomogeneity on the scale of less than 1~$\mu{m}$.
\end{abstract}
\maketitle

Graphene is the prototype two dimensional carbon
system~\cite{geimrev} and a promising candidate for future electronics\cite{kimribbon,Lemme,avouris}.
Graphene samples are usually obtained from micro-mechanical cleavage
of graphite~\cite{Novoselov2005proc} and they can be identified by elastic and inelastic light scattering, such as
Rayleigh and Raman spectroscopies \cite{ray,acf}.

Raman spectroscopy is a fast and non-destructive method for the
characterization of carbons\cite{acftrans}. Their Raman spectra shows common features in the 800-2000 cm$^{-1}$ region:
the G and D peaks, which lie at around 1560 and 1360
cm$^{-1}$ respectively. The G peak corresponds to the $E_{2g}$
phonon at the Brillouin zone centre. The D peak is due to
the breathing modes of sp$^2$ atoms and requires a defect for its activation\cite{tuinstra,Ferrari00,steffi}. It is common for as-prepared graphene not to have enough structural defects for the D peak to be Raman active\cite{acf}, so that it can only be seen at the edges\cite{acf}. However, the most prominent feature in
graphene is the second order of the D peak: the 2D peak\cite{acf}. This lies at $\sim$ 2700 cm$^{-1}$ and it is always seen, even when no D peak is present, since no defects are required for the activation of second order phonons. Its shape distinguishes single and multi-layer samples. Graphene has a sharp, single 2D peak, in contrast with graphite and few-layers graphene\cite{acf}.

The ability to controllably dope n or p is key for applications. Raman spectroscopy can monitor doping in graphene\cite{Pisana,das07}. The effect of back gating and top gating on the G-peak position (Pos(G)) and its Full Width at Half Maximum (FWHM(G)) was reported in Refs.~\cite{Pisana,YanPrl2007,das07}. Pos(G) increases and FWHM(G)
decreases for both electron and hole doping.
The stiffening of the G peak is due to the non-adiabatic removal of the
Kohn-anomaly at $\Gamma$\cite{LazzeriPrl06,Pisana}. The FWHM sharpening is due to blockage
of the phonon decay into electron-hole pairs due to the
Pauli exclusion principle, when the electron-hole gap becomes higher
than the phonon energy~\cite{Pisana,Lazzeri2006}. FWHM(G) sharpening saturates when doping causes a Fermi level
shift bigger than half the phonon energy\cite{Pisana,YanPrl2007}. A similar behavior is observed for the
LO-G$^{-}$ peak in metallic nanotubes\cite{Wu,shim}, for the same
reasons.

Much of previous research focussed on the properties of well defined graphene layers and devices\cite{geimrev,Novoselov2005proc,avouris,sciencetrans,naturekim,ray,acf,Pisana,das07,YanPrl2007}, with little effort
on a systematic investigation of sample variability.
Here we show that Raman spectroscopy can fingerprint differences between nominally identical samples,
produced in the same way. We find that, even in the absence of a D peak, changes in the Raman parameters are most common and relate to the presence of excess charges. This is a significant finding, which reconciles the variation of electrical properties often found for nominally identical samples.
\begin{figure}
\centerline{\includegraphics[width=50mm]{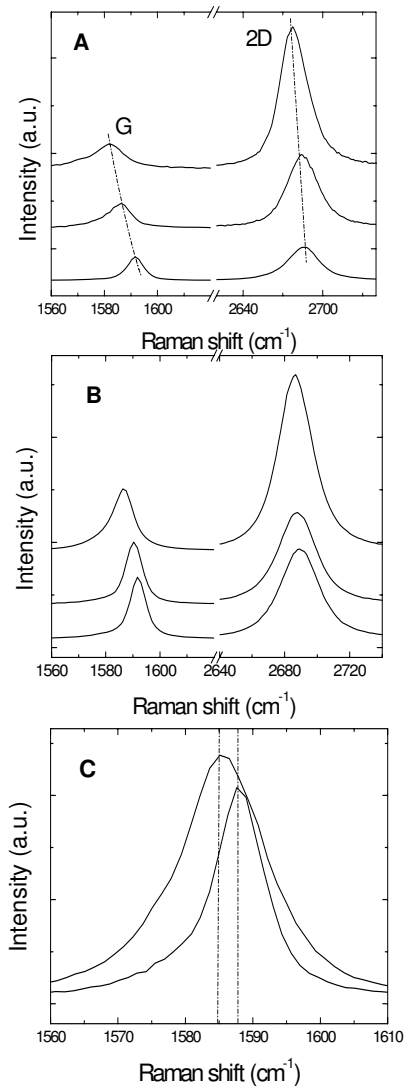}} \caption{(a)
514 nm spectra taken of three different
graphene samples; (b) spectra in three different points of the same sample. (c) The G peak can sometimes be asymmetric} \label{fig1}
\end{figure}
We study more than 40 as-prepared monolayer graphenes, produced by micro-cleavage of
graphite. These have different areas, from few $\mu$ m$^2$ to 450 $\mu$m$^2$. Some
of them are also measured in a device configuration, after deposition of Au electrodes
(with a thin Cr underlayer). More than $\sim$100 spectra are measured with a 100X objective at 514 and 633 nm with a Renishaw spectrometer, with  $\sim$ 2 cm$^{-1}$ spectral resolution and power well below 2 mW.

Fig. 1(a) plots the 514nm spectra of \textit{different} samples, normalized to the G peak. The G peak significantly shifts. The 2D
peak also shows a small change in position. The relative intensity of the 2D and G peaks strongly varies.  Fig 1(b) plots spectra measured on \textit{the same} graphene sample. This is a contacted sample, and the spectra change moving closer to the electrodes. Fig 1(c) indicates that the G peak can be sometimes asymmetric. Note that Fig. 1 does not mean that the Raman spectra always vary in different samples or that they always change within a given sample. However, it warns that uniformity has to be checked, and cannot be simply assumed. Moreover, Fig 1 dismisses the suggestion of Refs.\cite{GuptaNanoLett2006, GrafNanoLett2007} that either G peak position or I(2D)/I(G) can be used to estimate the number of layers, since the variation of these parameters in as deposited single layers far exceeds that assigned to the increase of number of layers\cite{GuptaNanoLett2006,GrafNanoLett2007}. Note that the criterium based on the shape of the 2D peak\cite{acf} still stands and allows layer counting.
\begin{figure}
\centerline{\includegraphics[width=55mm]{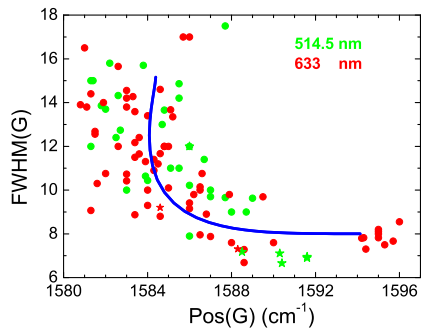}}
\caption{FWHM(G), Pos(G) at 514, 633 nm. Stars indicate samples
with metallic contacts. Only spectra without D peak are fitted.
The solid line is the theory for doped graphene at
300K\cite{Pisana}, giving more than 10$^{13}$ cm$^{-2}$ doping for
the bottom-right samples\cite{Pisana,das07}}\label{fig2}
\end{figure}
\begin{figure}
\centerline{\includegraphics[width=55mm]{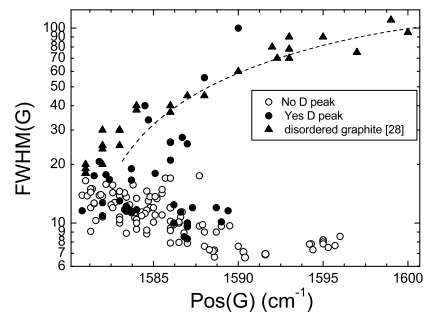}}
\caption{FWHM(G), Pos(G) for graphene with and without D peak, and
for nanocrystalline graphite\cite{lespade}}\label{fig2}
\end{figure}

Fig. 2 plots Pos(G) and FWHM(G). There
is a clear correlation: a Pos(G) increase corresponds to a FWHM(G)
decrease. This is quite similar to what we observed
in intentionally doped graphene, where the Fermi energy was
modulated using a gate~\cite{Pisana,das07}. Indeed, the continuous line
in Fig. 2 plots the theoretical correlation between Pos(G) and FWHM(G), obtained combining Eq.~(6,7) of Ref.\cite{Pisana}.
The agreement with experiments is remarkable, considering
that Ref.~\cite{Pisana} studied a single sample as a function
of doping, while Fig. 2 is a collection of measurements on tens of different samples, with no intentional control of doping. The star data-points in Fig 2 are measurements on contacted samples. Interestingly they usually have a significant doping. This is consistent chemical doping during micro-fabrication procedures, which can often be seen as a shift of the charge neutrality point away from zero gate voltage\cite{sciencetrans,Pisana,YanPrl2007,kimnew,puddles,das07}. However, it is quite remarkable that "pristine" samples, with no contacts, exhibit almost an order of magnitude doping variation, with a few showing a very high doping over $\sim10^{13}$cm$^{-2}$. Excess charges can be due to substrate, adsorbates, and resist/process residuals\cite{schedin}. In contacted samples, the difference of work function between sample and contacts can also contribute to the doping variation across the layer.

Fig 2 shows that the maximum FWHM(G) for the most intrinsic samples is $\sim$16 cm$^{-1}$, slightly higher than in graphite\cite{Lazzeri2006,Piscanec2004}. Note that all spectra used to derive Fig. 2 do not show any D peak. Thus, we exclude a significant influence of defects in the measured trend. Interestingly, as already observed in Refs.\cite{Pisana,YanPrl2007}, FWHM(G) never becomes smaller than $\sim$6 cm$^{-1}$, while for very high doping we would expect the minimum FWHM(G) to be close to our spectral resolution ($\sim$2 cm$^{-1}$). This implies an inhomogeneous distribution of charges within the laser spot of $\sim$ 1 $\mu$m$^2$ even for high self-doping. The asymmetric spectra of Fig. 1c indicate even larger variations.

Fig.3 includes data from samples with a D peak. Some fall in the same FWHM(G)/Pos(G) relation for D-peak-free samples, indicating that they originate from sample edges, not from disorder. However, others have FWHM(G) above 16 cm$^{-1}$, the maximum measured for D-peak-free samples, accompanied by a stiffening of the G peak. This is a signature of structural disorder\cite{Ferrari00,lespade,SSCReview}. Indeed, in the case of graphite, it is known that, for increasing defects leading to nano-crystalline graphite, FWHM(G) and Pos(G) both increase\cite{Ferrari00,lespade,SSCReview}, the opposite of what happens for increasing doping. Thus, a large FWHM(G), together with Pos(G) close to 1580 cm$^{-1}$ and no D peak fingerprint the most intrinsic samples, while a large FWHM(G), Pos(G) higher than 1580cm$^{-1}$ and a D peak indicate structural disorder.

\begin{figure}
\centerline{\includegraphics[width=50mm]{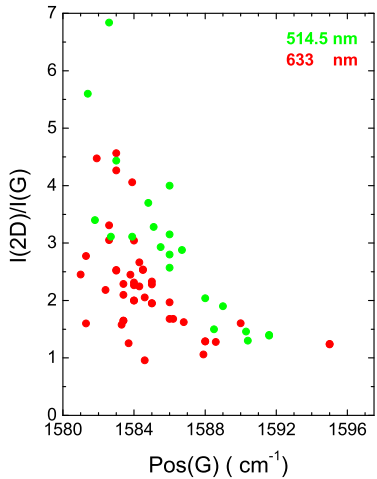}}
\caption{I(2D)/I(G) as a function of Pos(G).} \label{fig4}
\end{figure}
\begin{figure}
\centerline{\includegraphics[width=55mm]{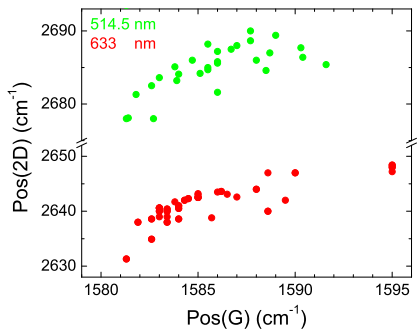}}
\caption{Pos(2D) as a function Pos(G) at 514 and 633 nm} \label{fig3}
\end{figure}

We now analyze the 2D peak. Fig. 4 plots I(2D)/I(G) as a function of Pos(G). This clearly shows a large variation with doping: at low doping the 2D peak is 3-5 times stronger than the G peak, depending on the excitation wavelength; at high doping (for a G peak position above 1592 cm$^{-1}$) the intensity ratio is $\sim$1.

Figure 5 correlates Pos(2D) and Pos(G). Unlike the G peak, the 2D peak always upshifts with excitation
energy, due to double resonance~\cite{steffi,acf}. The dispersion with excitation energy is $95-85 ~cm^{-1}$/eV. Fig. 5 also shows that the 2D peak is sensitive to doping. Doping has two major effects: (i) modification of the equilibrium lattice parameter,
with a consequent stiffening/softening of the phonons~\cite{Pietronero};(ii) onset of dynamic effects beyond the Born-Oppenheimer approximation that modify the phonon
dispersions close to the Kohn anomalies~\cite{Pisana,LazzeriPrl06}. For the 2D peak
the influence of dynamic effects is expected to be
negligible, since the 2D phonons are far
away from the Kohn Anomaly at K\cite{acf,das07,Piscanec2004}. Thus, the variation of the 2D peak with
doping is mainly due to charge transfer, with hole doping resulting in an upshift, and the opposite for high electron doping\cite{das07}. Indeed, FWHM(2D) does not show the same trend as FWHM(G), but is $\sim$ 28-30 cm$^{-1}$ for all samples. Since Fig. 5 indicates 2D stiffening with increasing Pos(G), we conclude that most of our samples show hole doping. This agrees with what found in electrical measurements, where the charge neutrality points is mostly reached for positive gate bias\cite{sciencetrans,das07}. Adsorbents induce chemical
doping and water could explain the p-doping\cite{schedin}.

In conclusion, we presented a systematic analysis of the Raman spectra of
as-deposited graphene. When no D peak is present, the large variation in Raman parameters is assigned to charged impurities. Variations in the Raman spectra can also be observed within the same sample, indicating in-homogeneous charges. A D peak far from the edge means structural disorder. Thus, Raman is a powerful tool to monitor the "quality" of graphene.

CC acknowledges the Oppenheimer Fund. ACF, AKG, KSN the Royal Society and
Leverhulme Trust.

\end{document}